\documentclass[twocolumn]{aastex62}

\newcommand\Ha{H$\alpha$}

\newcommand\hii{{\sc H\thinspace ii}}

\newcommand\civ{{\sc C\thinspace iv}}
\newcommand\oiii{{\sc O\thinspace iii}}
\newcommand\oii{[{\sc O\thinspace ii}]}
\newcommand\ovi{{\sc O\thinspace vi}}
\newcommand\heii{He{\sc \thinspace ii}}
\newcommand\kms{{\rm\,km\,s^{-1}}}
\newcommand\ergs{{\rm\,erg\,s^{-1}}}
\newcommand\zsol{\rm\,Z_\odot}
\newcommand\msol{\rm\,M_\odot}

\newcommand\mrk{{Mrk~71}}

\graphicspath{{./}{figures/}}

\received{September 1, 2023}
\revised{October 17, 2023}
\accepted{October 18, 2023}
\submitjournal{ApJ Letters}

\shorttitle{Nebular C IV Imaging of Mrk 71}
\shortauthors{Oey et al.}

\begin{document}

\title{NEBULAR C IV $\lambda1550$ IMAGING OF THE METAL-POOR STARBURST MRK 71: \\
DIRECT EVIDENCE OF CATASTROPHIC COOLING}


\correspondingauthor{M. S. Oey}
\email{msoey@umich.edu}

\author[0000-0000-0000-0000]{M. S. Oey}
\affil{University of Michigan,
Department of Astronomy, 1085 South University Ave.,
Ann Arbor, MI 48109-1107, USA}

\author[0000-0000-0000-0000]{Amit N. Sawant}
\affil{University of Michigan,
Department of Astronomy, 1085 South University Ave.,
Ann Arbor, MI 48109-1107, USA}
\affil{Present address:  
Institute of Mathematics,
EPFL SB MATH,
MA C2 647,
Station 8,
CH-1015 Lausanne, Switzerland}

\author{Ashkbiz Danehkar}
\affiliation{Eureka Scientific,
2452 Delmer Street, Suite 100, Oakland, CA 94602-3017, USA}

\author{Sergiy Silich}
\affiliation{Instituto Nacional de Astrof\'isica, \'Optica y Electr'onica (INAOE),
AP 51, 72000 Puebla, M\'exico}

\author{Linda J. Smith}
\affiliation{Space Telescope Science Institute,
3700 San Martin Dr.,
Baltimore, MD 21218, USA}

\author{Jens Melinder}
\affiliation{Stockholm University,
Department of Astronomy and Oskar Klein Centre,
AlbaNova University Centre,
SE-10691 Stockholm, Sweden}

\author{Claus Leitherer}
\affiliation{Space Telescope Science Institute,
3700 San Martin Dr.,
Baltimore, MD 21218, USA}

\author{Matthew Hayes}
\affiliation{Stockholm University,
Department of Astronomy and Oskar Klein Centre,
AlbaNova University Centre,
SE-10691 Stockholm, Sweden}

\author{Anne E. Jaskot}
\affiliation{Williams College,
Department of Astronomy,
Williamstown, MA 01267, USA}

\author{Daniela Calzetti}
\affiliation{
Department of Astronomy,
University of Massachusetts,
Amherst, MA   01003, USA}

\author{You-Hua Chu}
\affiliation{Institute of Astronomy and Astrophysics, Academia Sinica (ASIAA),
No. 1, Sec. 4, Roosevelt Road, 
Taipei 10617, Taiwan}

\author{Bethan L. James}
\affiliation{Space Telescope Science Institute,
3700 San Martin Dr.,
Baltimore, MD 21218, USA}

\author{G\"oran \"Ostlin}
\affiliation{Stockholm University,
Department of Astronomy and Oskar Klein Centre,
AlbaNova University Centre,
SE-10691 Stockholm, Sweden}
 
\begin{abstract}
We use the Hubble Space Telescope ACS camera to obtain the first spatially resolved, nebular imaging in the light of \civ\ $\lambda\lambda1548,1551$ by using the F150LP and F165LP filters.  These observations of the local starburst Mrk~71 in NGC~2366 show emission apparently originating within the interior cavity around the dominant super star cluster (SSC), Knot~A. Together with imaging in \heii\ $\lambda4686$ and supporting STIS FUV spectroscopy, the morphology 
and intensity of the \civ\ nebular surface brightness and the \civ/\heii\ ratio map provide direct evidence that the mechanical feedback is likely
dominated by catastrophic radiative cooling, which strongly disrupts adiabatic superbubble evolution.  The implied extreme mass loading and low kinetic efficiency of the cluster wind are reasonably consistent with the wind energy budget, which is probably enhanced by radiation pressure.
In contrast, the Knot~B SSC lies within a well-defined superbubble with associated soft X-rays and \heii\ $\lambda1640$ emission, which are signatures of adiabatic, energy-driven feedback from a supernova-driven outflow.  
This system lacks clear evidence of \civ\ from the limb-brightened shell, as expected for this model, but the observations may not  be deep enough to confirm its presence.
We also detect 
a small \civ-emitting object that is likely an embedded compact \hii\ region.  Its \civ\ emission may indicate the presence of very massive stars ($> 100\ \msol$) or strongly pressure-confined stellar feedback.
\end{abstract}

\keywords{starburst galaxies --- galaxy winds --- galaxy evolution --- emission-line galaxies --- stellar feedback --- young massive clusters --- superbubbles --- \hii\ regions --- ultraviolet photometry --- direct imaging}

\section{Introduction} \label{sec:intro}

Massive star feedback encompasses energetic processes that heat gas to temperatures above $10^4$ K. OB stars and their hot, blue descendants, as well as high-mass X-ray binaries, photoionize gas into this regime, and shock-heating by supernovae and stellar winds drive temperatures up to $10^6$ to $10^8$ K.  
\civ\ $\lambda\lambda 1548,1551$ (hereafter ``\civ\ $\lambda1550$") 
is ubiquitous in the interstellar medium of star-forming galaxies \citep[e.g.,][]{Savage1984, Savage2001, Wang2005}, where it is believed to originate in conductive interfaces between hot ($> 10^6$) K gas and cooler ISM phases \citep[e.g.,][]{McCray1979}; this
species is also a prominent P-Cygni emission line arising in hot star winds.
\civ\ $\lambda1550$ traces ionization energies above 47.9 eV, and for recombination, above 64.5 eV.    
Nebular \civ\ is therefore only rarely seen, and is associated more with planetary nebulae rather than ordinary \hii\ regions 
\citep[e.g.,][]{Aller1981, Harrington1982}.

However, \civ\ $\lambda1550$ emission does appear in a number of extreme starbursts, both locally \citep{Mingozzi2022, Berg2019, Senchyna2019} and at high redshift, where it can be prominent \citep{Stark2015, Amorin2017, Senchyna2022}.  In these objects, it is generally thought to be nebular, although its origin is not well understood.  It could be a signature of photoionization by unusually hot stars like Wolf-Rayet (WR) stars or rapidly rotating stars, or by high-mass X-ray binaries (HMXBs); or it could be due to mechanical feedback, whether from direct collisional ionization and conductive interfaces to adiabatic heating zones \citep[e.g.,][]{Chu1994}, or from radiative, catastrophic cooling flows \citep{Gray2019a} that disrupt adiabatic conditions.   
The origin of \civ\ emission is of particular cosmological interest \citep{Senchyna2022} when linked to low metallicites.  This nebular line implies higher ionization parameters than is normally seen in \hii\ regions, and different mechanisms have been proposed to explain its presence in intense metal-poor starbursts.  Under these conditions stars are more compact, with faster stellar rotation, both of which increase the effective temperatures.  Low metallicity is also linked to stronger interaction in close binaries, which promotes the formation of WR stars and fast rotators by binary mass transfer, as well as the creation of HMXBs.

Nebular \civ\ imaging of resolved, local objects would therefore provide an important and revealing diagnostic for these different scenarios.  But since these $\lambda\lambda1548,1551$ resonance lines are in the far ultraviolet, it has generally been inaccessible for such targets.  However, the Solar Blind Channel (SBC) of the Advanced Camera for Surveys (ACS) aboard the Hubble Space Telescope (HST) offers a long-pass filter set, F150LP and F165LP, that is capable of imaging in the light of \civ\ $\lambda1550$.  The net transmission in these filters is shown by \citet{Hayes2016}, who used a similar filter pair to successfully carry out imaging of the starburst galaxy SDSS J115630.63+500822.1 in \ovi\ $\lambda\lambda1032,1038$.  In this Letter, we report the first spatially resolved imaging of \civ\ nebular emission, which was carried out with the ACS/SBC.

\begin{figure*}[ht!]
\plotone{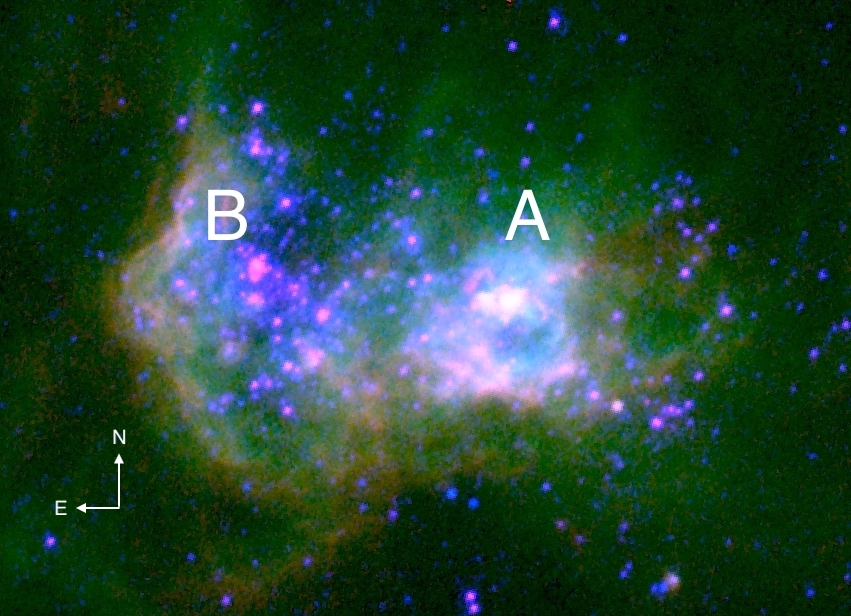}
\smallskip
\vspace*{5pt}
\plotone{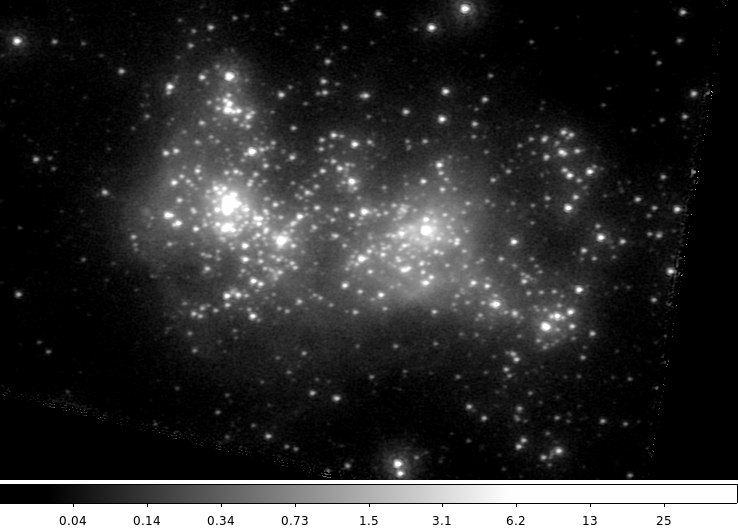}
\caption{
Top: Three-color HST/WFC3 image of \mrk\ in F373N (\oii\ $\lambda$3727), F502N ([\oiii] $\lambda$5007), and F469N (He {\sc ii} $\lambda$4686) corresponding to red, green, and blue, respectively.  These archive data were obtained by \citet{James2016} (GO-13041; colors not to scale). Bottom: Our new, total combined F150LP image, in units of $10^{-16}\ \rm erg\ s^{-1}\ cm^{-2}$, not corrected for reddening.  Most of the diffuse emission seen in this image is due to scattered starlight.  Knots A and B are separated by 5.0$\arcsec$ (83 pc).
\label{f_mrk71}}
\end{figure*}

Our target is the local starburst complex Mrk 71 in the nearby Magellanic irregular galaxy NGC 2366.  This system is of intense interest since it is a remarkable analog of extreme Green Pea galaxies \citep{Micheva2017}, which are the only known class of local Lyman-continuum emitting galaxies \citep[e.g.,][]{Izotov2018a, Flury2022a}. 
At a distance of only 3.4 Mpc \citep{Tolstoy1995}, \mrk\ is close enough to resolve individual stars.
It is also a metal-poor system, having $12+\log(\rm O/H)=7.89$ \citep{Izotov1997, Chen2023}, or about $0.16\zsol$.  The nature of feedback changes dramatically at this low metallicity.  In addition to hotter stellar photoionizing sources described above, mechanical feedback is much weaker \citep{Jecmen2023} since at low metallicity, supernovae occur mostly at the lower-mass range of core collapse progenitors \citep{Patton2020, OConnor2011, Heger2003} and stellar winds are dramatically weaker \citep[e.g.,][]{Vink2022,Ramachandran2019,Bjorklund2023}.  Mrk 71 provides an outstanding template for metal-poor feedback processes because it hosts both a young super star cluster (SSC) driving an extreme ionization parameter (Knot A), and a second, more evolved, SSC (Knot B) that has generated a mature superbubble system (Figure~\ref{f_mrk71}).  Our \civ\ imaging yields critical, diagnostic insight on both of these subsystems.

\section{\civ\ imaging observations}

We obtained Cycle 28 ACS/SBC imaging observations of \mrk\ (GO-16261; PI Oey) in F150LP and F165LP during 2020 Oct 28 -- Nov 01, using the LODARK aperture.
The target was observed in F150LP for 4 $\times 1480$ s plus 4 $\times 1497$ s, yielding a total of 11,908 s.  In F165LP, the total exposure was $8\times 1480$ s plus $8 \times 1497$ s plus 3054 s, or 26,870 s total.  We combined the frames in each of the two bands using the STScI DrizzlePac software.  The world coordinate systems of the images were first aligned to an accuracy of 0.15 -- 0.20 pixels, as limited by the non-gaussian point-spread function \citep[PSF; see][]{Avila2016}. 
We then drizzled the 8 F150LP frames together, 
obtaining a pixel scale of $0\arcsec.025$ pixel$^{-1}$.
The 17 F165LP images were drizzled as a separate set, using the combined, drizzled F150LP image as the reference.  PSF matching was carried out using the Photutils package, convolving both drizzled images with a combined PSF.
The final combined F150LP image is shown in the bottom panel of Figure~\ref{f_mrk71}.  Extensive, diffuse emission is seen throughout the region; however, most of this extended emission corresponds to scattered starlight as noted by \citet{Drissen2000}, and we show below that it is removed by continuum subtraction.  

\begin{figure}
\plotone{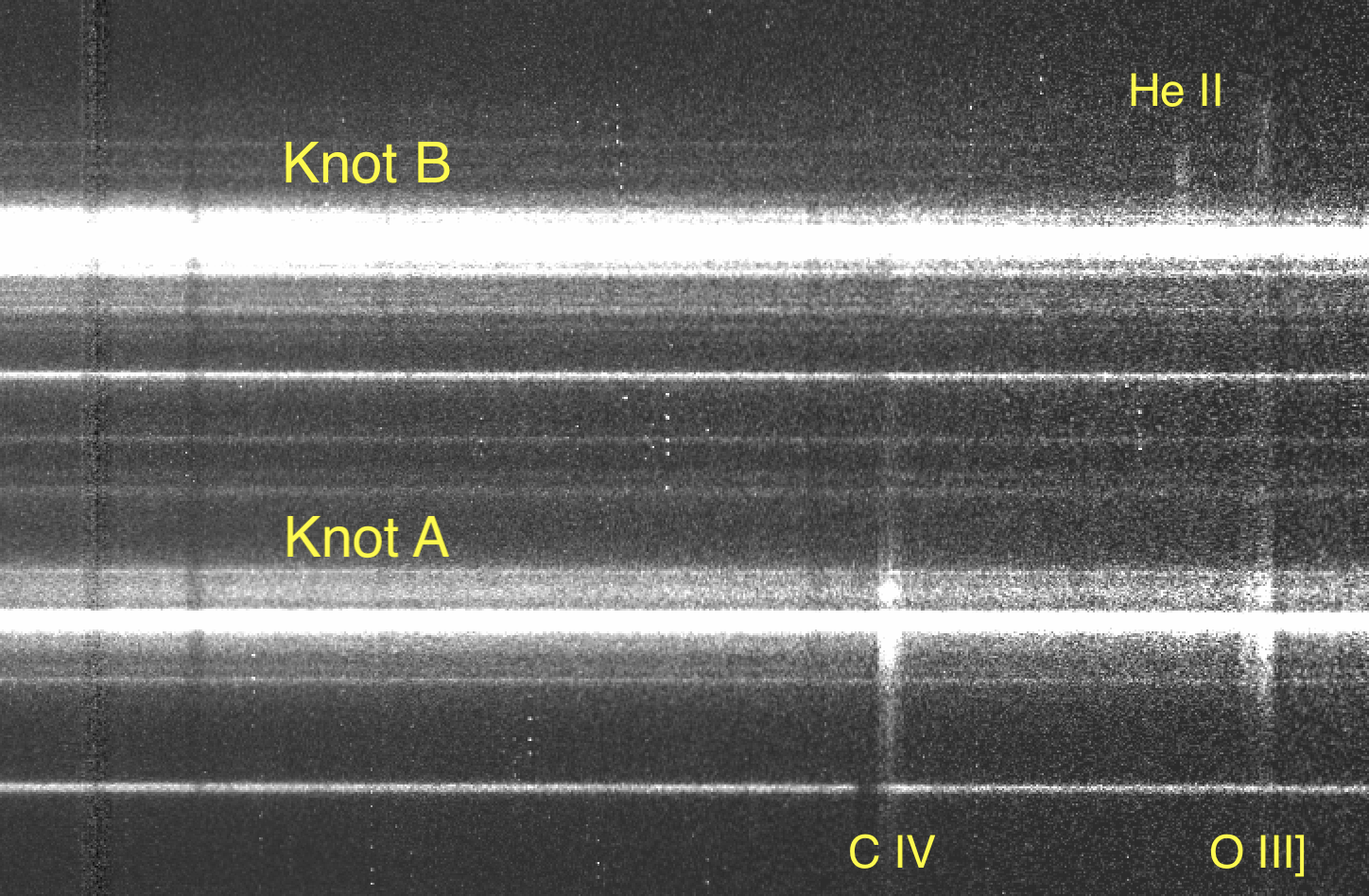}
\caption{
STIS FUV long-slit spectrum across Knots A and B.  The nebular emission features of \civ\ $\lambda1550$, \heii\ $\lambda 1640$ and {\sc Oiii]} $\lambda\lambda 1661, 1666$ are marked.  
\label{f_STIS}}
\end{figure}

\begin{figure*}
\plotone{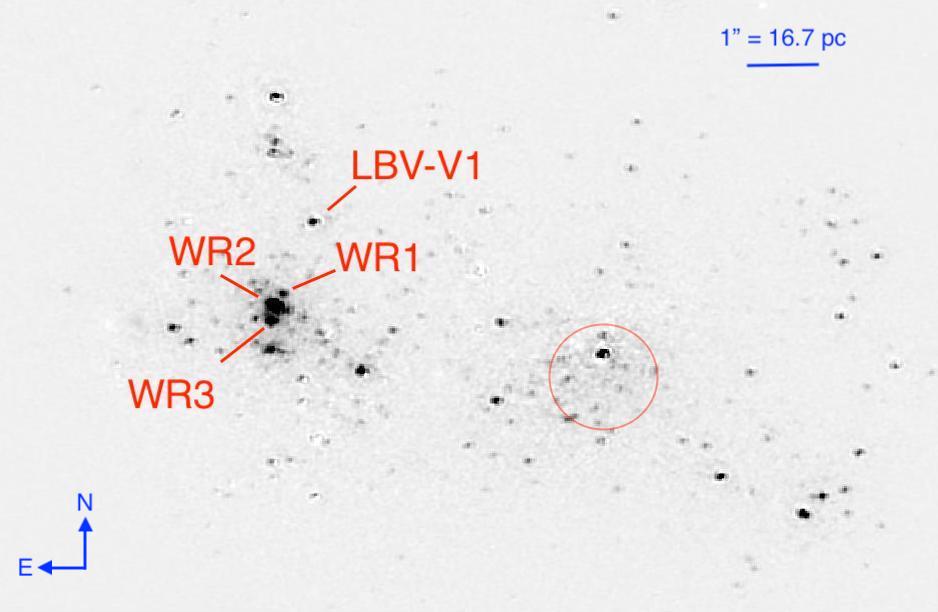}
\smallskip
\vspace*{5pt}
\plotone{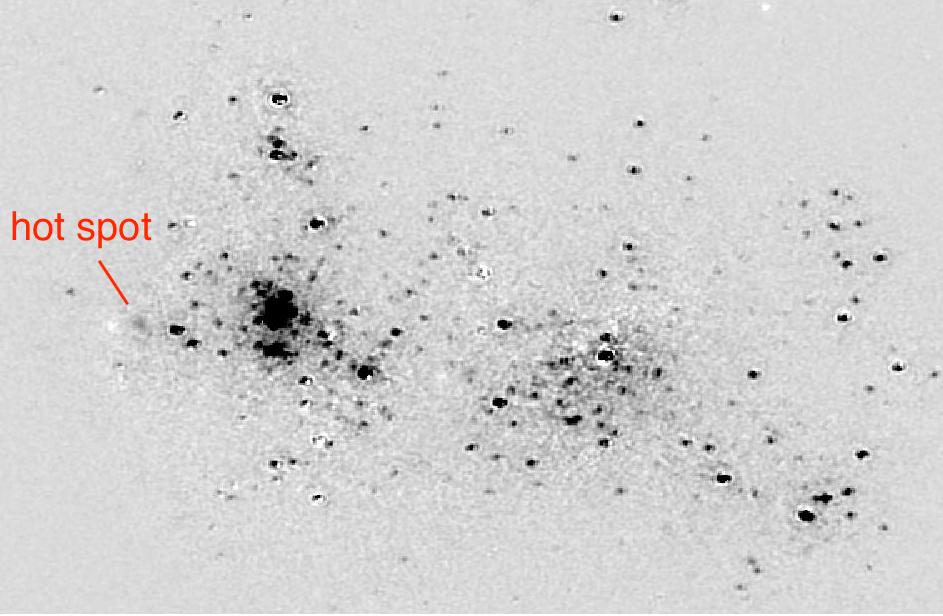}
\caption{The top and bottom panels 
show shallow and deep logarithmic contrasts, respectively, of the continuum-subtracted \civ\ $\lambda\lambda1548,1551$ image.  WR stars, LBV-V1, and an IR-bright ``hot spot'' \citep{Drissen1997,Drissen2000} are identified. 
The circle around Knot A indicates the $0.8\arcsec$-radius aperture used to measure the \civ\ nebular flux.
\label{f_civ_image}}
\end{figure*}

Our HST program also obtained a deep STIS long-slit spectrum across Knots~A and B, using the G140L grating which includes the \civ\ spectral region (Figure~\ref{f_STIS}).  These data are reduced and calibrated in our study of the stellar population in Knot~A \citep{Smith2023}.  We use this spectrum to determine the scale factor for the F165LP image to match the depth in F150LP at \civ, thereby flux-calibrating the continuum-subtracted \civ\ image.  This is carried out by matching the observed diffuse, 
{\bf nebular}
\civ\ flux in two regions with $0.\arcsec5$ length along the $0.\arcsec 2$-wide slit on both sides adjacent to Knot~A.  The resulting scale factor
is consistent with the results obtained using the method developed by \citet{Sawant2021} based on maximizing the modal bin fraction.
Shallow and deep versions of the final continuum-subtracted image are shown in Figure~\ref{f_civ_image}.  
We note that the \heii\ $\lambda1640$ emission line falls within the lower throughput regimes of both the F150LP and F165LP filters; the transmission curves and F165LP scale factor imply that the level of remaining \heii\ $\lambda1640$ emission in the continuum-subtracted image is on the order of 2\% of the signal.  Below, we specifically compare to archive \heii\ $\lambda4686$ emission.

Many stars are seen in the deep panel of Figure~\ref{f_civ_image}. 
Since, as described above, the continuum subtraction is applied to the \civ\ emission and is independent of the stellar data, these stars are mostly strong FUV emitters, and are not necessarily \civ\ emitters.  This is discussed further in Section~\ref{s_KnotB}.  
The detection of stars is also enhanced by our deep continuum imaging, which generates r.m.s. noise in the background of only $\sim 0.1$\%.
Diffuse emission is also apparently seen around both Knots~A and B in Figure~\ref{f_civ_image}.  However, inspection of the STIS spectrum in Figure~\ref{f_STIS} shows that only the diffuse emission around Knot~A is real.
We believe that the spurious emission around Knot B is caused by the SBC PSF.  While its FWHM is $0.02\arcsec$ and its 50\% encircled energy radius is 0.075$\arcsec$ \citep{Avila2016}, it has broad, faint wings that appear to be imperfectly continuum subtracted and thus may cause residual emission around the brightest stars.  The effect extends out to about $\sim0\arcsec.25$, as can be seen around, e.g., LBV-V1  (Figure~\ref{f_civ_image}).  Knot B is 4.6$\times$ brighter than Knot A within an aperture of 0.35$\arcsec$ radius, and it also has many more UV-bright stars in a more spatially extended configuration, whereas Knot A does not have many such stars in its immediate vicinity. 

\section{Knot A:  Catastrophic Cooling} \label{s_KnotA}

The Knot~A SSC strongly dominates the luminosity \citep[e.g.,][]{Gonzalez-Delgado1994, Drissen2000} and Green Pea-like ionization parameter \citep[$\log U = -2$;][]{James2016} of the \mrk\ complex.  Its estimated mass is 
$\sim 1.4\times 10^5\ \msol$ based on the \Ha\ luminosity \citep{Micheva2017} 
and its age is $1\pm 1$ Myr based on stellar population synthesis \citep{Smith2023}.
The SSC is still substantially enshrouded, but stellar features implying the presence of VMS stars are detected \citep{Smith2023}.

Nebular \civ\ $\lambda1550$ emission can originate from systems with weak mechanical feedback via strong radiative cooling and/or high-energy photoionization of dense gas retained near the parent SSC. Or, \civ\ can be emitted from a conductive interface between a cool, dense shell and interior hot gas generated by strong, energy-driven mechanical feedback.  Thus, these two scenarios produce contrasting morphologies in 
nebular \civ: interior emission for weak mechanical feedback versus shell emission for strong mechanical feedback \citep[e.g.,][]{Danehkar2022, Danehkar2021, Gray2019a}, as demonstrated further below.

\begin{figure*}
\plotone{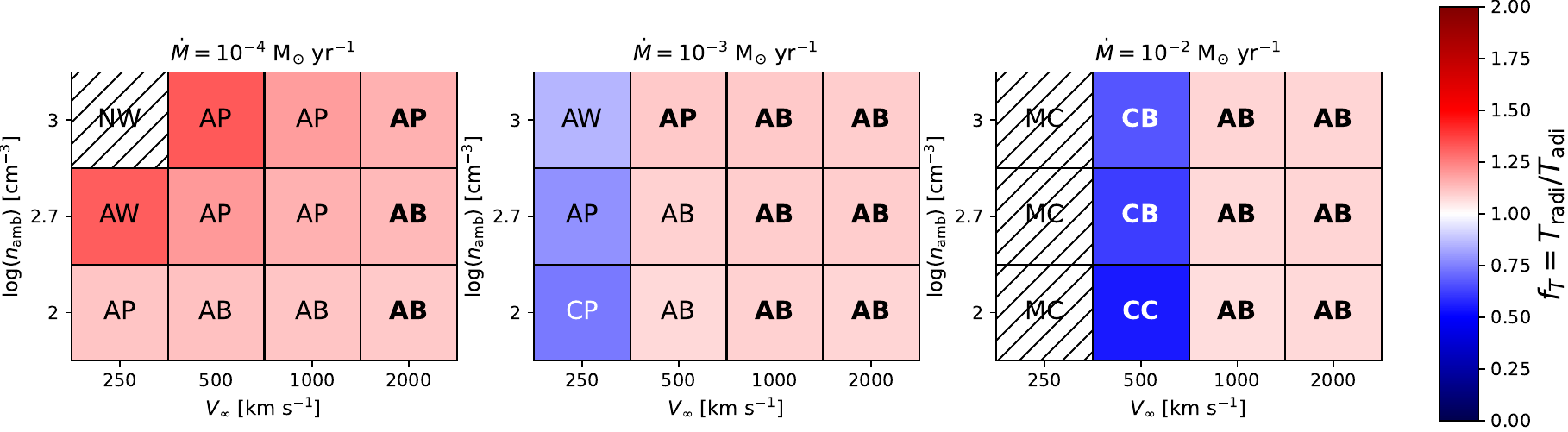}
\caption{
Outcomes for the modeled parameter space following \citet[][see text]{Danehkar2021}, where AB = adiabatic bubble, AP = pressure-confined adiabatic bubble, AW = adiabatic wind only, NW = no wind, 
CB = partially cooling bubble}, 
CC = full catastrophic cooling, CP = cooling pressure-confined, MC = momentum conserving only.  The code fails for the NW and MC cases since these do not generate energy-driven models.  The color scale shows the temperature of the hot bubble region (Figures~\ref{f_models1} and \ref{f_models2}) relative to the value expected for simple adiabatic expansion. See \citet{Danehkar2021} for a full explanation and discussion of these categories.
\label{f_grid}
\end{figure*}

Figure~\ref{f_civ_image} shows diffuse \civ\ emission within a $sim 15$-pc radius of the Knot~A SSC, assuming a distance of 3.4 Mpc \citep{Tolstoy1995}.  This region is coincident with the boundary of the dense gas to the west and south (Figure~\ref{f_mrk71}) that has been identified as a cavity or shell created by mechanical feedback from the SSC \citep{Komarova2021a, Oey2017}.  
Observations of the nebular and molecular gas kinematics for this region show that the shell expansion velocity is only 
$\sim 5 -  10\ \kms$ 
\citep{Komarova2021a, Micheva2019, Oey2017}.  
This localized offset in the systemic velocity is a separate component from the faint, broad emission-line wings that are a much more spatially extended feature discussed in detail by \citet{Komarova2021a}.
For the observed parameters associated with Knot~A, this local shell expansion velocity is consistent with momentum-conserving, non-adiabatic expansion and thus
this system has been suggested \citep{Komarova2021a, Oey2017} to be an example of metal-poor feedback where superwinds are suppressed \citep[e.g.,][]{Jecmen2023}.

Weak, dense superwinds can experience 
strong, radiative
cooling that quenches the energy-driven, adiabatic outflow \citep[e.g.,][]{Silich2004,  Krumholz2009, Lochhaas2021}, and furthermore, weak winds may be suppressed within the cluster itself, such that the individual stellar wind bubbles fail to merge into a coherent outflow \citep{Silich2018a, Yadav2017}.  
In the last case, an expanding cavity is formed by photoionization and radiation pressure from the SSC.  We refer to both of these scenarios 
that disrupt adiabatic evolution as ``catastrophic" cooling.  Following \citet{Danehkar2021}, we adopt a criterion that the driving superwind has dropped to a temperature $<75\%$ of the adiabatic value.
Another possibility is that energy-driven feedback may exist but may be pressure-confined \citep{Oey2004b, Silich2007}.  
Thus, we note that while the shell velocity and parameter space for this object are more suggestive of momentum-conserving evolution \citep[e.g.,][]{Komarova2021a}, the kinematics alone are insufficient to distinguish between an energy-driven, adiabatic and non-adiabatic, catastrophic cooling regime.  This underscores the importance of ions like \civ\ as a diagnostic of the interior and shell temperature structure.

\begin{figure*}
\epsscale{1.12}
\plotone{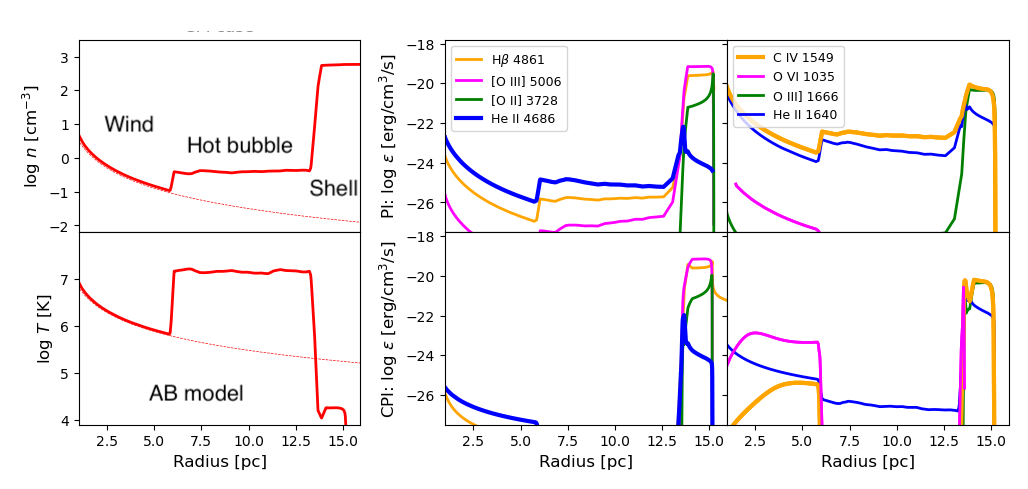}
\caption{
Models of volume emissivity ($\rm erg\ s^{-1}\ cm^{-3}$) for the shown emission lines as a function of radius, for 
a conventional adiabatic model (AB) generated with $V = 1000\ \kms$, $\log\dot{M} = -3 ~/ \rm \msol\ yr^{-1}$, and $n = 1000\ \rm cm^{-3}$ from the grid in Figure~\ref{f_grid}.
We also show the radial profiles for the temperature and density of this model in the left panels.  The dotted lines correspond to the analytic relations $n \sim r^{-2}$ and $T \sim r^{-4/3}$  for an adiabatic, freely expanding wind.
The bottom emissivity panels show calculations for combined collisional and photoionization (CPI) that are predicted for this model. As a comparison, the top panels show models for pure photoionization (PI) of the same density distribution, which would be largely isothermal on the order of $10^4$ K \citep{Danehkar2021}.
\label{f_models1}}
\end{figure*}

\begin{figure*}
\epsscale{1.12}
\plotone{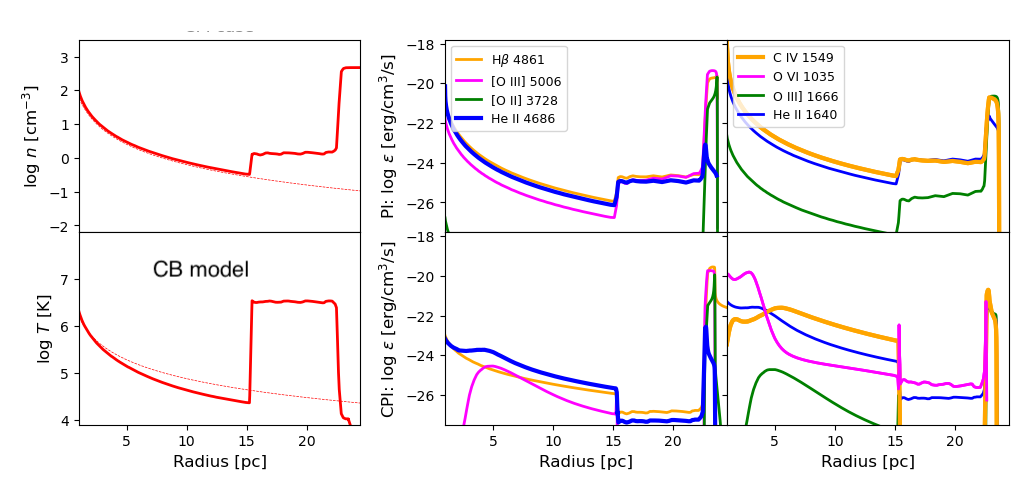}
\caption{
The same as Figure~\ref{f_models1}, but for a
catastrophic cooling model (CB) with 
$V = 500\ \kms$, $\log\dot{M} = -2 ~/ \rm \msol\ yr^{-1}$, and $n = 500\ \rm cm^{-3}$.
The hot bubble interior is now much cooler than for the AB model in Figure~\ref{f_models1} and occupies a much smaller fractional volume interior to the shell.
\label{f_models2}}
\end{figure*}

Following \citet{Danehkar2021}, we calculate a grid of models for energy-driven feedback
using the {\sc Maihem} non-equilibrium ionization code \citep{Gray2019a}, covering a range of parameters similar to those inferred for the Knot~A SSC.  We assume a cluster radius of 1~pc \citep{Micheva2017}, mass $1\times 10^5\ \msol$ with a Salpeter IMF having stellar mass range $0.5 - 150\ \msol$, age 1~Myr, and metallicity $0.1\ \zsol$. The effective wind velocity is modeled in the range $V = 250 - 2000\ \kms$, its effective mass-loss rate is in the range $\log\dot{M} = -2$ to $-4 ~/~ \rm \msol\ yr^{-1}$, and the ambient density is $n = 100 - 1000\ \rm cm^{-3}$.  
The modeled ranges for $V$ and $\dot{M}$ are based on those expected for the combined stellar winds from the SSC and extend to values that generate suppressed superwind conditions similar to those in \citep{Danehkar2021}.
Figure~\ref{f_grid} shows the distribution of feedback outcomes, ranging from fully adiabatic bubbles to fully momentum-conserving outflows.

In Figures~\ref{f_models1} and \ref{f_models2}, we show radial emissivities calculated following \citet{Danehkar2022} using {\sc Cloudy} \citep{Ferland2017} 
for two of the grid models that are broadly similar to the observed shell size, and which offer contrasting interpretations of adiabatic  and catastrophic cooling, respectively.
The model generated with $V = 1000\ \kms$, $\log\dot{M} = -3 / \rm \msol\ yr^{-1}$, and $n = 1000\ \rm cm^{-3}$ produces a conventional adiabatic bubble (AB; Figure~\ref{f_models1}); 
and the other is a strongly cooling model, with 
$V = 500\ \kms$, $\log\dot{M} = -2 ~/ \rm \msol\ yr^{-1}$, and $n = 500\ \rm cm^{-3}$ 
(CB; Figure~\ref{f_models2}).  
The density and temperature profiles for these models are also shown, and
the radial zones corresponding to the freely expanding SSC wind with the given $\dot{M}$ and $V$, the shock-heated hot bubble, and the dense outer shell are indicated on 
the density profile in Figure~\ref{f_models1}.  Figure~\ref{f_models2}
demonstrates that the CB model's temperature is below the adiabatic prediction in the expanding wind region.  Thus, its hot bubble temperature is several times lower, on the order of a few $\times 10^6$ K, than that of the AB model, which has $T > 10^7$ K.  We also see that the CB hot bubble region occupies a much lower fractional volume interior to the shell.  For stronger cooling, the hot region is cooler and shrinks further.

The emissivity calculations in the bottom panels include both kinetic and photoionizing activation in the evolving outflow.  We see that for the fully adiabatic model, essentially all the \civ\ $\lambda1550$ emission comes from the shell, and virtually none from the interior (Figure~\ref{f_models1}).  In contrast, for the strongly cooling model, the interior contributes substantally to the \civ\ emission over a large volume \citep[Figure~\ref{f_models2};][]{Gray2019a, Danehkar2022}.

The top panels in Figures~\ref{f_models1} and \ref{f_models2} show emissivities of only photoionization for the same density distribution.  
This model is largely isothermal at $T\sim 10^4$ K, and demonstrates
the difference when excluding kinetic heating.  These are useful for inferring the emissivity profiles for pure momentum-conserving evolution (MC), where there is no shock heating.  In that case, the line emission should generally follow the $r^{-2}$ profiles in the inner radial zone, which corresponds to the driving wind's density profile; for complete catastrophic cooling, i.e., pure MC evolution, this zone would extend directly to the shell, with no hot bubble zone.

To measure the nebular \civ\ flux around Knot~A, we spatially interpolate over the stars within the circular emitting region of radius $0.\arcsec8$ (13 pc), obtaining a total observed flux of 
$3.33\pm 0.50\times10^{-14}\ \rm \ergs\ cm^{-2}$.
We apply the foreground $E(B-V) = 0.033$ using the Milky Way extinction law \citep{Cardelli1989} and local $E(B-V) = 0.084$ for Knot~A following \citet{Smith2023}, using the SMC reddening from \citet{Gordon2003}.  This yields a dereddened flux of 
$6.84\pm 1.03 \times10^{-14}\ \rm \ergs\ cm^{-2}$
and surface brightness of 
$3.41\pm 0.51\times 10^{-14}\ \rm \ergs\ cm^{-2}\ arcsec^{-2}$.
The uncertainties account for measurement errors only; we estimate that systematic uncertainties due to continuum subtraction, stellar interpolation, and reddening correction are on the order of 50\%, 10\%, and 30\%, respectively.  We also caution that there is uncertainty due to the imperfect PSF subtraction described above, but it is difficult to quantify and Knot~A does not have many UV-bright stars in its immediate vicinity relative to the 0.25$\arcsec$ limit at which the PSF effect is seen.

Resolved surface brightness can be written as $3.74\times10^{-12} \epsilon_i\ dr$ \citep{Hazy2013}, where $\epsilon_i$ is the \civ\ $\lambda 1550$ emissivity and $dr$ is the line-of-sight path length.
A value of $\epsilon_i\sim 1\times 10^{-22}\ \rm erg\ s^{-1}\ cm^{-3}$ (see Figure~\ref{f_models2}) integrated through 20 pc yields a surface brightness $\sim 2\times 10^{-14}\ \ergs\ \rm cm^{-2}\ arcsec^{-2}$, which agrees well with the observed values.  Given that the predicted values are for the relatively crude approximations taken from our model grid, this serves to demonstrate good general consistency between expectations and observations.

\begin{figure}
\plotone{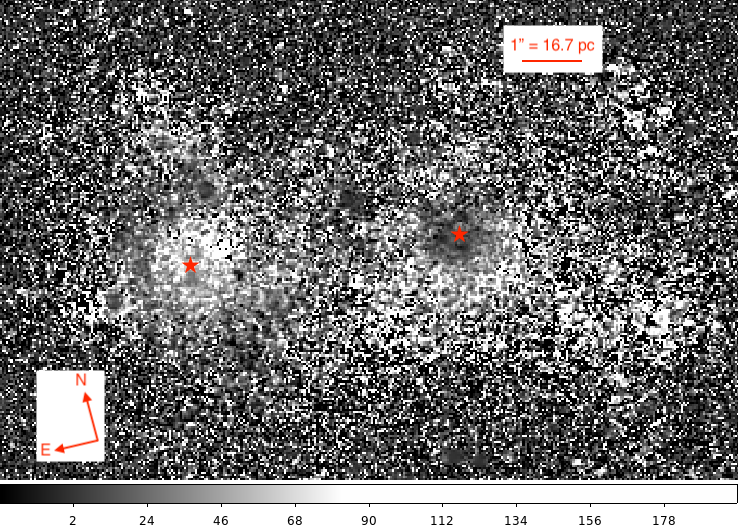}
\caption{ 
Ratio map of continuum-subtracted \civ\ $\lambda1550$ / \heii\ $\lambda4686$, with the locations of Knots A and B marked.  The central enhancement around Knot~B is an artifact (see Figure~\ref{f_STIS} and Section~\ref{s_KnotB}).
\label{f_C4He2}}
\end{figure}

Figure~\ref{f_civ_image} shows that the \civ\ nebular emission around Knot~A appears more consistent with an internally emitting morphology than a limb-brightened, dense shell.  We can also compare with the \heii~$\lambda4686$ morphology, which, similar to \civ, has only shell emission for strong feedback and interior emission for weak, radiatively cooling feedback (bottom panels of Figures~\ref{f_models1} and \ref{f_models2}, respectively). 
Figure~\ref{f_C4He2} shows the ratio map of continuum-subtracted, nebular \civ\ $\lambda1550$ / \heii\ $\lambda4686$ using the archive image in F469N from \citet[][]{James2016}.  
These data show that \civ/\heii\ is lower in the center around Knot~A and higher at larger radii.  The centrally concentrated morphology of \heii\ was noted by \citet{James2016}, and it is not apparent in \civ\ (Figure~\ref{f_civ_image}).  
The region with reduced \civ/\heii\ extends to a radius $> 0.5\arcsec$, which is much larger than the barely resolved $\sim 0.1\arcsec$ Knot A.  The \heii\ imaging was obtained with the WFC3 camera, which does not have the PSF wing issue described above for the ACS SBC MAMA detector, and so the flux cannot be attributed to the \heii-emitting stars \citep{Smith2023} in the core.

\begin{figure}
\plotone{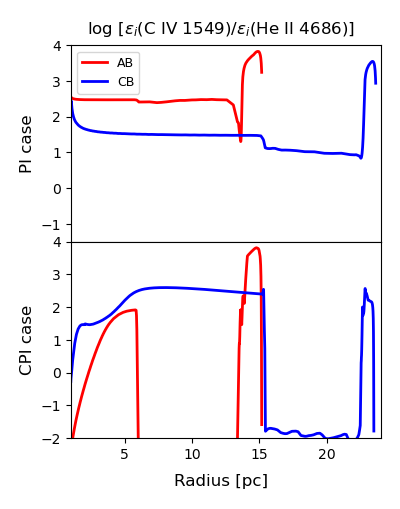}
\caption{
Predicted emissivity ratio of \civ\ to \heii\ as a function of radius for the AB (red) and CB (blue) models shown in Figures~\ref{f_models1} and \ref{f_models2}, respectively.  Pure photoionization models are shown in the top panel and combined collisional $+$ photoionization are shown in the bottom panels.  We caution that the emissivities of these lines are orders of magnitude lower in the AB model, as seen in Figure~\ref{f_models1}.
\label{f_Ratiomodels}}
\end{figure}

As seen in Figure~\ref{f_Ratiomodels}, this pattern seen in the ratio map is consistent with that expected for the strongly cooling CB model.
The radial morphology of both \civ\ and \heii\ in the CB model is based on the original $r^{-2}$ wind density profile seen in the upper panels of Figures~\ref{f_models1} and \ref{f_models2} for pure photoionization, which is then modified by kinetic heating such that the central zones become more highly ionized \citep{Gray2019a}, depressing the emissivities for these species in the center, as seen in the bottom panels of the figures.  They also show enhanced emission at larger radii.  However, the two ions have slightly differing radial profiles, resulting in the radial dependence of their ratio shown in Figure~\ref{f_Ratiomodels} for the CB model (blue), and which is seen in our data (Figure~\ref{f_C4He2}).

\citet{Smith2023} determine the likely presence of very massive stars (VMS) from our STIS spectrum.  These may have masses up to $600 \msol$, and thus
may contribute to the high ionization implied by \civ\ and \heii\ emission \citep[e.g.,][]{Berg2019}.
In contrast, for the classical AB model, the emissivity ratio for \civ/\heii\ is essentially zero for most of the bubble volume (Figure~\ref{f_Ratiomodels}), and the interior emissivities,
including in the central wind region, are orders of magnitude lower and undetectable (Figure~\ref{f_models1}). 

For the CB model in Figure~\ref{f_models2}, \heii\ $\lambda1640$ is brighter than \civ\ $\lambda1550$ at the smallest radii.  However, the STIS spectrum (Figure~\ref{f_STIS}) shows no detection of \heii\ $\lambda1640$, even near the SSC itself.
This does not rule out catastrophic cooling, since the central \heii\ $\lambda1640$ is not always stronger than \civ\ $\lambda1550$ for such models \citep{Danehkar2022, Gray2019a}.  On the other hand,
the pure photoionization models (top panels of Figures~\ref{f_models1} and \ref{f_models2}) have  \civ\ $\lambda1550$ / \heii\ $\lambda1640 > 1$ in the entirety of the central wind zone.  This may suggest that pure photoionization dominates here for Knot~A.  This is supported by the fact that \oiii] $\lambda\lambda1661, 1666$ is also seen in this region.
While \oiii] is seen across the entire Knot~A environment, much of this flux is likely to be foreground and background emission since the entire object is enveloped in a large halo of more diffuse, extended [\oiii] (Figure~\ref{f_mrk71}); however, the central enhancement is coincident with the strongest \civ\ emission around the SSC (Figure~\ref{f_STIS}) 
and suggests local \oiii] emission from Knot~A itself.
It is important to note that the models are idealized and remnant high-density gas is known to be in the immediate vicinity of the SSC, which remains largely enshrouded, at least along the line of sight \citep[e.g.,][]{Smith2023, Micheva2017}.  Pure photoionization is consistent with complete cooling of any energy-driven, mechanical feedback.

Overall, the detection of significant diffuse, nebular \civ\ emitted from gas within an SSC-generated cavity must be linked to the absence of pure energy-driven feedback.  In such a scenario, the \civ\ emission originates either from catastrophic radiative cooling that disrupts adiabatic feedback, or from pure photoionization; both scenarios are incompatible with the shock-heated temperatures ($\gtrsim 10^6$ K) required for the classical, adiabatic model, which would ionize C to higher levels.  These conclusions are therefore robust to parameters like shell geometry and metallicity variations.  

The Knot A system has parameters that are quantitatively very similar to those of the embedded SSC and surrounding molecular gas of NGC 5253 D-1 \citep[e.g.,][]{Turner2017}.  In particular,
the molecular velocity dispersion of $\sim 10\ \kms$ and gas mass $\sim 10^5\ \msol$ observed to be within a few pc of Knot~A \citep{Oey2017} are similar to values for NGC 5253 D-1, which \citep{Silich2023} show are consistent with radiative, catastrophic cooling conditions.
The catastrophic cooling models develop in our grid for $\log\dot{M} = -2\ / \rm\msol\ yr^{-1}$ and $V \leq 500\ \kms$. 
This is quantitatively consistent with the models for NGC 5253-D1 by \citet{Silich2020}, who
find that a mass deposition rate on the order of $\dot{M}\sim 10^{-2}\ \rm\msol\ yr^{-1}$ prevents individual stellar winds from merging and developing a global outflow.  
This would also be linked to a strongly cooling environment.

This effective $\dot{M}$ is roughly 2 orders of magnitude higher than what is expected for a $10^5\ \msol$ SSC, and $V$ may be up to an order of magnitude lower than expected for stellar wind velocities on the order of 2000 $\kms$.  These factors are quantitatively consistent with conservation of the SSC wind's kinetic energy:  a factor of 100 increase in mass balances a factor of 10 reduction in wind velocity to conserve total kinetic energy.
This suggests that extreme mass-loading of the wind is directly linked to its deceleration or kinetic inefficiency.  The most straightforward process would be by ablation and shredding of material from molecular gas clouds and clumps,
which are known to be in close proximity to the SSC and interior to the shell, at least in the line of sight \citep{Oey2017}.  Disks of pre-main sequence stars are also suggested as a source of material for mass loading \citep{Silich2020}. 

However, metal-poor stellar winds may have lower velocities, on the order of 1000 $\kms$ \citep[e.g.,][]{Garcia2014}.  Thus, $V$ may be as little as a factor of 2, instead of an order of magnitude, below that expected for the SSC.  It is therefore likely that mass-loading is enhanced by another process, in particular, by
photoevaporation, which is expected in this radiation-dominated environment that includes large quantities of molecular gas.  
Moreover, radiation is dynamically important, as indicated by the extreme ionization parameter \citep{Dopita2002, Yeh2012} associated with Knot~A, and by the fact that radiation is also believed to concurrently drive a separate, very low-density, fast superwind through openings in the shell walls \citep{Komarova2021a}.
Radiation therefore likely contributes to maintaining the outward momentum of the slow, mass-loaded flow associated with the shell around Knot~A.

\section{Knot B and Other Features}\label{s_KnotB}

Although its mass is up to 10$\times$ lower than Knot~A \citep{Micheva2017, Drissen2000}, the Knot~B SSC is by far the brightest source in Figure~\ref{f_civ_image}.  The emission from this cluster is dominated by 3 known WR stars identified by \citet{Drissen2000}, including WR3, an unusually luminous ($M_V \sim -7$) WC star.
We caution that although many stars are clearly detected in the continuum-subtracted image, they are not necessarily \civ-emitting stars.  As described in Section~\ref{s_KnotA}, the continuum subtraction is calibrated to the diffuse, nebular \civ\ emission.  This continuum normalization is not optimized to identify stellar \civ\ emission, since most such stars have extremely blue continuum slopes and the long-pass filter bandpasses are very broad.  Thus the point sources in Figure~\ref{f_civ_image} are mostly very FUV-bright stars.  Some of the brightest objects in the figure may be \civ-emitters, but without spectroscopic confirmation, they should only be considered candidates.
These include the luminous blue variable LBV-V1 \citep{Drissen1997, Petit2006}; P-Cygni \civ\ emission has sometimes been observed in other LBVs, \citep[e.g., HD 5980;][]{Koenigsberger1995}.
We will carry out detailed analysis of the stellar population, including its FUV properties, in a follow-up paper.

Figures~\ref{f_mrk71}, \ref{f_civ_image}, and \ref{f_C4He2} appear to suggest the existence of substantial diffuse \civ\ emission near the central SSC in Knot~B.  
As noted earlier, this emission is not seen in the STIS spectrum (Figure~\ref{f_STIS}) and is therefore a spurious effect likely linked to the large, faint PSF wings of the MAMA detector and 
the more complex, extended morphology of this SSC, with numerous FUV-bright stars in the vicinity of the central core, consistent with its older age. 
Thus, the appearance of a high \civ/\heii\ ratio around Knot B in Figure~\ref{f_C4He2} is an artifact.  The F469N image obtained with WFC3 does not have broad PSF wings and therefore does not suffer from this effect.

The large, dense shell around Knot B (Figure~\ref{f_mrk71}) is likely due to the action of multiple supernovae from this somewhat older \citep[$\gtrsim 4$ Myr;][]{Micheva2017} SSC.  
This is supported by the existence of diffuse, soft X-ray emission associated with the southern region of this shell \citep{Thuan2014}, which is a clear signature of adiabatic, energy-driven feedback.

The lower panels of Figure~\ref{f_models1} show the typical relative emissivities for AB models.  While the radial ranges of the different wind and hot bubble zones may vary depending on the specific parameters, the radial morphologies with the three shown zones are all similar for strong, adiabatic feedback.  In particular, \civ\ is ordinarily suppressed near the bubble center due to the prevalence of {\sc C v} \citep[e.g.,][]{Gray2019a}.  The lack of central, diffuse \civ\ around Knot~B is therefore fully consistent with adiabatic feedback.  Furthermore, the presence of central \heii\ $\lambda1640$ seen in the STIS spectrum (Figure~\ref{f_STIS}) is also consistent with the AB model in Figure~\ref{f_models1}.  As noted earlier, the diffuse \oiii] emission in the spectrum is likely foreground or background emission.

However, there is no apparent detection of limb-brightened \civ\ from the shell wall, as expected from the cooling interface with the hot, X-ray emitting gas \citep[e.g.,][]{Chu1994, Danehkar2022}.  
Figure~\ref{f_models1} shows that for an AB model, the shell's predicted emissivity is on the order of $10^{21}\ \rm erg\ s^{-1}\ cm^{-3}$; thus for a $\sim 3$-pc line of sight through the limb, the surface brightness might be around $3\times 10^{-14}\ \rm erg\ s^{-1}\ cm^{-2}\ arcsec^{-2}$, which is similar to that in the cavity around Knot~A.  
It may be that the emission is a bit fainter than expected and the observations are not deep enough to detect the emission.

A non-stellar, dusty ``hot spot'' that is bright in $K$-band \citep{Drissen2000} also appears to be detected in nebular \civ\ emission (Figure~\ref{f_civ_image}).  It is compact, with a FWHM 
$\lesssim 0.15\arcsec$, 
or 2.5 pc,
and the colors reported by \citet{Drissen2000} of $J - H = 0.4$ and $H - K = 1.5$ are consistent with those of compact \hii\ regions \citep[e.g.,][]{Kastner2008}.
The total \civ\ flux of the object is
$> 2.81\pm 0.08\times 10^{-15}\ \ergs\ \rm cm^{-2}$, taking the extinction correction used for Knot~A.  
The quoted uncertainty includes only measurement error, while systematic uncertainty due to continuum subtraction is on the order of 10\%.  However,
this is a substantial lower limit, since the extinction for such a dense object is unknown and expected to be much higher than for Knot~A.  
As noted earlier, \civ\ emission from \hii\ regions is unusual and implies the presence of higher-energy photons.  
Assuming the object is a compact \hii\ region, it may host one or more extremely hot, early O stars, perhaps VMS stars; alternatively, the \civ\ emission may be generated in the conductive interface of a very compact shell enclosing a hot bubble that is strongly pressure-confined \citep[e.g.,][]{Silich2007, Danehkar2021}.  Such a system may also generate \civ\ by photoionization from the hot gas \citep{Oskinova2022}.
The location of this object suggests that its formation may be triggered by the shell expansion due to Knot~B.

\section{Conclusion}

Our successful use of the F150LP and F165LP filters to obtain imaging in \civ\ $\lambda 1550$ demonstrates the viability of the ACS SBC channel for obtaining spatially resolved nebular observations in this critical diagnostic emission line. Extended diffuse emission is seen in the \mrk\ system, and also a highly excited, compact  \hii\ region.  
FUV-bright stars, including some candidate stellar \civ\ sources like WR stars, early O stars, and LBV-V1 are clearly detected in our observations, although further analysis is needed to determine the extent to which \civ-emitting stars can be distinguished from stars with steep FUV continua. 

For the Knot~A system, 
provided that
the observed diffuse \civ\ emission is real, as confirmed by the STIS spectrum, its center-filled spatial distribution and lack of limb-brightened morphology imply that mechanical feedback is non-adiabatic, as previously suggested by the surrounding shell kinematics.
Moreover, its morphology and observed flux is consistent with model expectations for strong radiative cooling.
The observed lower \civ/\heii\ ratio in the center is also characteristic of such cooling.
It may be a system that is not quite a completely cooled, purely momentum-conserving system, and/or there may be total cooling and pure photoionization, especially within Knot~A itself.
These observations therefore provide {\it direct evidence} that Knot~A is 
likely
driving a momentum-conserving shell due to catastrophic cooling that disrupts adiabatic, energy-driven kinematics. 

The Knot~B SSC has generated a well-defined superbubble associated with diffuse X-rays.  There is no perceptible \civ\ emission from within the superbubble, while central \heii\ $\lambda1640$ is detected, as expected from the SSC wind.  These factors all point to a system dominated by adiabatic, supernova-driven feedback.  
However, there is no immediate evidence of limb-brightened \civ\ emission from the superbubble shell that is predicted for adiabatic models.  Deeper observations are needed to establish the extent to which this represents a significant discrepancy with predictions.

\acknowledgments

We thank Will Gray, Genoveva Micheva, and Megan Reiter for useful discussions, 
and Roberto Avila of the STScI Astrodrizzle team for advice on the drizzle procedure.  
We also thank the anonymous referee for useful comments and questions.
This work was supported by NASA HST-GO-16261.  S.S. is supported by CONAHCYT, M\'exico, research grant A1-S-28458. This research is based on observations made with the NASA/ESA Hubble Space Telescope obtained from the Space Telescope Science Institute, which is operated by the Association of Universities for Research in Astronomy, Inc., under NASA contract NAS 5–26555. These observations are associated with program 16261.  
The data presented in this paper were obtained from the Mikulski Archive for Space Telescopes (MAST) at the Space Telescope Science Institute. The specific observations analyzed can be accessed via \dataset[DOI: 10.17909/90na-ch15]{https://doi.org/10.17909/90na-ch15}.


\vspace*{5mm}
\facilities{HST(ACS, STIS)}

\software{DrizzlePac\citep{Hoffman2021}}

\bibliographystyle{aasjournal}{}
\bibliography{Oey_Mrk71_CIV}

\begin{thebibliography}{}
\expandafter\ifx\csname natexlab\endcsname\relax\def\natexlab#1{#1}\fi
\providecommand{\url}[1]{\href{#1}{#1}}

\bibitem[{{Aller} {et~al.}(1981){Aller}, {Ross}, {Omara}, \&
  {Keyes}}]{Aller1981}
{Aller}, L.~H., {Ross}, J.~E., {Omara}, B.~J., \& {Keyes}, C.~D. 1981, \mnras,
  197, 95

\bibitem[{Amorín {et~al.}(2017)Amorín, Fontana, Pérez-Montero, Castellano,
  Guaita, Grazian, Fèvre, Ribeiro, Schaerer, Tasca, Thomas, Bardelli,
  Cassarà, Cassata, Cimatti, Contini, Barros, Garilli, Giavalisco, Hathi,
  Koekemoer, Le~Brun, Lemaux, MacCagni, Pentericci, Pforr, Talia, Tresse,
  Vanzella, Vergani, Zamorani, Zucca, \& Merlin}]{Amorin2017}
Amorín, R., Fontana, A., Pérez-Montero, E., {et~al.} 2017, Nature Astronomy,
  1, 0057

\bibitem[{Avila \& Chiaberge(2016)}]{Avila2016}
Avila, R.~J., \& Chiaberge, M. 2016, Photometric {Aperture} {Corrections} for
  the {ACS}/{SBC}, Tech. Rep. ACS 2016-05, STScI

\bibitem[{{Berg} {et~al.}(2019){Berg}, {Chisholm}, {Erb}, {Pogge}, {Henry}, \&
  {Olivier}}]{Berg2019}
{Berg}, D.~A., {Chisholm}, J., {Erb}, D.~K., {et~al.} 2019, \apjl, 878, L3

\bibitem[{{Bj{\"o}rklund} {et~al.}(2023){Bj{\"o}rklund}, {Sundqvist}, {Singh},
  {Puls}, \& {Najarro}}]{Bjorklund2023}
{Bj{\"o}rklund}, R., {Sundqvist}, J.~O., {Singh}, S.~M., {Puls}, J., \&
  {Najarro}, F. 2023, \aap, 676, A109

\bibitem[{{Cardelli} {et~al.}(1989){Cardelli}, {Clayton}, \&
  {Mathis}}]{Cardelli1989}
{Cardelli}, J.~A., {Clayton}, G.~C., \& {Mathis}, J.~S. 1989, \apj, 345, 245

\bibitem[{{Chen} {et~al.}(2023){Chen}, {Jones}, {Sanders}, {Fadda}, {Sutter},
  {Minchin}, {Huntzinger}, {Senchyna}, {Stark}, {Spilker}, {Weiner}, \&
  {Roberts-Borsani}}]{Chen2023}
{Chen}, Y., {Jones}, T., {Sanders}, R., {et~al.} 2023, Nature Astronomy, 7, 771

\bibitem[{{Chu} {et~al.}(1994){Chu}, {Wakker}, {Mac Low}, \&
  {Garcia-Segura}}]{Chu1994}
{Chu}, Y.~H., {Wakker}, B., {Mac Low}, M.~M., \& {Garcia-Segura}, G. 1994, \aj,
  108, 1696

\bibitem[{Danehkar {et~al.}(2021)Danehkar, Oey, \& Gray}]{Danehkar2021}
Danehkar, A., Oey, M.~S., \& Gray, W.~J. 2021, ApJ, 921, 91.
\newblock \url{https://ui.adsabs.harvard.edu/abs/2021ApJ...921...91D}

\bibitem[{Danehkar {et~al.}(2022)Danehkar, Oey, \& Gray}]{Danehkar2022}
---. 2022, ApJ, 937, 68.
\newblock \url{https://iopscience.iop.org/article/10.3847/1538-4357/ac8cec}

\bibitem[{{Dopita} {et~al.}(2002){Dopita}, {Groves}, {Sutherland}, {Binette},
  \& {Cecil}}]{Dopita2002}
{Dopita}, M.~A., {Groves}, B.~A., {Sutherland}, R.~S., {Binette}, L., \&
  {Cecil}, G. 2002, \apj, 572, 753

\bibitem[{Drissen {et~al.}(2000)Drissen, Roy, Robert, \& Devost}]{Drissen2000}
Drissen, L., Roy, J.-L., Robert, C., \& Devost, D. 2000, AJ, 119, 688

\bibitem[{{Drissen} {et~al.}(1997){Drissen}, {Roy}, \& {Robert}}]{Drissen1997}
{Drissen}, L., {Roy}, J.-R., \& {Robert}, C. 1997, \apjl, 474, L35

\bibitem[{Ferland(2013)}]{Hazy2013}
Ferland, G.~J., e.~a. 2013, Hazy: A Brief Introduction to Cloudy C13.1, Tech.
  rep., Cloudy \& Associates.
\newblock \url{http://www.nublado.org}

\bibitem[{{Ferland} {et~al.}(2017){Ferland}, {Chatzikos}, {Guzm{\'a}n},
  {Lykins}, {van Hoof}, {Williams}, {Abel}, {Badnell}, {Keenan}, {Porter}, \&
  {Stancil}}]{Ferland2017}
{Ferland}, G.~J., {Chatzikos}, M., {Guzm{\'a}n}, F., {et~al.} 2017, \rmxaa, 53,
  385

\bibitem[{{Flury} {et~al.}(2022){Flury}, {Jaskot}, {Ferguson}, {Worseck},
  {Makan}, {Chisholm}, {Saldana-Lopez}, {Schaerer}, {McCandliss}, {Wang},
  {Ford}, {Heckman}, {Ji}, {Giavalisco}, {Amorin}, {Atek}, {Blaizot},
  {Borthakur}, {Carr}, {Castellano}, {Cristiani}, {De Barros}, {Dickinson},
  {Finkelstein}, {Fleming}, {Fontanot}, {Garel}, {Grazian}, {Hayes}, {Henry},
  {Mauerhofer}, {Micheva}, {Oey}, {Ostlin}, {Papovich}, {Pentericci},
  {Ravindranath}, {Rosdahl}, {Rutkowski}, {Santini}, {Scarlata}, {Teplitz},
  {Thuan}, {Trebitsch}, {Vanzella}, {Verhamme}, \& {Xu}}]{Flury2022a}
{Flury}, S.~R., {Jaskot}, A.~E., {Ferguson}, H.~C., {et~al.} 2022, \apjs, 260,
  1

\bibitem[{{Garcia} {et~al.}(2014){Garcia}, {Herrero}, {Najarro}, {Lennon}, \&
  {Alejandro Urbaneja}}]{Garcia2014}
{Garcia}, M., {Herrero}, A., {Najarro}, F., {Lennon}, D.~J., \& {Alejandro
  Urbaneja}, M. 2014, \apj, 788, 64

\bibitem[{Gonz\'alez-Delgado {et~al.}(1994)Gonz\'alez-Delgado, Perez,
  Tenorio-Tagle, Vilchez, Terlevich, Terlevich, Telles, Rodriguez-Espinosa,
  Mas-Hesse, Garcia-Vargas, Diaz, Cepa, \& Castaneda}]{Gonzalez-Delgado1994}
Gonz\'alez-Delgado, R.~M., Perez, E., Tenorio-Tagle, G., {et~al.} 1994, ApJ,
  437, 239.
\newblock \url{http://adsabs.harvard.edu/doi/10.1086/174992}

\bibitem[{{Gordon} {et~al.}(2003){Gordon}, {Clayton}, {Misselt}, {Landolt}, \&
  {Wolff}}]{Gordon2003}
{Gordon}, K.~D., {Clayton}, G.~C., {Misselt}, K.~A., {Landolt}, A.~U., \&
  {Wolff}, M.~J. 2003, \apj, 594, 279

\bibitem[{Gray {et~al.}(2019)Gray, Oey, Silich, \& Scannapieco}]{Gray2019a}
Gray, W.~J., Oey, M.~S., Silich, S., \& Scannapieco, E. 2019, ApJ, 887, 161.
\newblock \url{http://dx.doi.org/10.3847/1538-4357/ab510d}

\bibitem[{{Harrington} {et~al.}(1982){Harrington}, {Seaton}, {Adams}, \&
  {Lutz}}]{Harrington1982}
{Harrington}, J.~P., {Seaton}, M.~J., {Adams}, S., \& {Lutz}, J.~H. 1982,
  \mnras, 199, 517

\bibitem[{Hayes {et~al.}(2016)Hayes, Melinder, Östlin, Scarlata, Lehnert, \&
  Mannerström-Jansson}]{Hayes2016}
Hayes, M., Melinder, J., Östlin, G., {et~al.} 2016, ApJ, 828, 1.
\newblock \url{http://dx.doi.org/10.3847/0004-637X/828/1/49}

\bibitem[{Heger {et~al.}(2003)Heger, Fryer, Woosley, Langer, \&
  Hartmann}]{Heger2003}
Heger, A., Fryer, C.~L., Woosley, S.~E., Langer, N., \& Hartmann, D.~H. 2003,
  ApJ, 591, 288

\bibitem[{{Hoffmann} {et~al.}(2021){Hoffmann}, {Mack}, \& {et
  al.}}]{Hoffman2021}
{Hoffmann}, S.~L., {Mack}, J., \& {et al.} 2021, The DrizzlePac Handbook,
  Version 2.0, Tech. rep., STScI

\bibitem[{Izotov {et~al.}(2018)Izotov, Schaerer, Worseck, Guseva, Thuan,
  Verhamme, Orlitová, \& Fricke}]{Izotov2018a}
Izotov, Y.~I., Schaerer, D., Worseck, G., {et~al.} 2018, \mnras, 474, 4514

\bibitem[{{Izotov} {et~al.}(1997){Izotov}, {Thuan}, \&
  {Lipovetsky}}]{Izotov1997}
{Izotov}, Y.~I., {Thuan}, T.~X., \& {Lipovetsky}, V.~A. 1997, \apjs, 108, 1

\bibitem[{James {et~al.}(2016)James, Auger, Aloisi, Calzetti, \&
  Kewley}]{James2016}
James, B.~L., Auger, M., Aloisi, A., Calzetti, D., \& Kewley, L. 2016, ApJ,
  816, 40.
\newblock
  \url{http://stacks.iop.org/0004-637X/816/i=1/a=40?key=crossref.22d4ebc9b775137b36b90e4a15b77786}

\bibitem[{Jecmen \& Oey(2023)}]{Jecmen2023}
Jecmen, M.~E., \& Oey, M.~S. 2023, ApJ, submitted

\bibitem[{{Kastner} {et~al.}(2008){Kastner}, {Thorndike}, {Romanczyk},
  {Buchanan}, {Hrivnak}, {Sahai}, \& {Egan}}]{Kastner2008}
{Kastner}, J.~H., {Thorndike}, S.~L., {Romanczyk}, P.~A., {et~al.} 2008, \aj,
  136, 1221

\bibitem[{{Koenigsberger} {et~al.}(1995){Koenigsberger}, {Guinan}, {Auer}, \&
  {Georgiev}}]{Koenigsberger1995}
{Koenigsberger}, G., {Guinan}, E., {Auer}, L., \& {Georgiev}, L. 1995, \apjl,
  452, L107

\bibitem[{{Komarova} {et~al.}(2021){Komarova}, {Oey}, {Krumholz}, {Silich},
  {Kumari}, \& {James}}]{Komarova2021a}
{Komarova}, L., {Oey}, M.~S., {Krumholz}, M.~R., {et~al.} 2021, \apjl, 920, L46

\bibitem[{Krumholz \& Matzner(2009)}]{Krumholz2009}
Krumholz, M.~R., \& Matzner, C.~D. 2009, ApJ, 703, 1352.
\newblock
  \url{http://stacks.iop.org/0004-637X/703/i=2/a=1352?key=crossref.1fe93aec04a09475adce4e6bfbe637b5}

\bibitem[{{Lochhaas} {et~al.}(2021){Lochhaas}, {Thompson}, \&
  {Schneider}}]{Lochhaas2021}
{Lochhaas}, C., {Thompson}, T.~A., \& {Schneider}, E.~E. 2021, \mnras, 504,
  3412

\bibitem[{{McCray} \& {Snow}(1979)}]{McCray1979}
{McCray}, R., \& {Snow}, T.~P., J. 1979, \araa, 17, 213

\bibitem[{Micheva {et~al.}(2019)Micheva, Christian~Herenz, Roth, Östlin, \&
  Girichidis}]{Micheva2019}
Micheva, G., Christian~Herenz, E., Roth, M.~M., Östlin, G., \& Girichidis, P.
  2019, A\&A, 623, 1

\bibitem[{Micheva {et~al.}(2017)Micheva, Oey, Jaskot, \& James}]{Micheva2017}
Micheva, G., Oey, M.~S., Jaskot, A.~E., \& James, B.~L. 2017, ApJ, 845, 165.
\newblock \url{http://dx.doi.org/10.3847/1538-4357/aa830b}

\bibitem[{{Mingozzi} {et~al.}(2022){Mingozzi}, {James}, {Arellano-C{\'o}rdova},
  {Berg}, {Senchyna}, {Chisholm}, {Brinchmann}, {Aloisi}, {Amor{\'\i}n},
  {Charlot}, {Feltre}, {Hayes}, {Heckman}, {Henry}, {Hernandez}, {Kumari},
  {Leitherer}, {Llerena}, {Martin}, {Nanayakkara}, {Ravindranath}, {Skillman},
  {Sugahara}, {Wofford}, \& {Xu}}]{Mingozzi2022}
{Mingozzi}, M., {James}, B.~L., {Arellano-C{\'o}rdova}, K.~Z., {et~al.} 2022,
  \apj, 939, 110

\bibitem[{O'Connor \& Ott(2011)}]{OConnor2011}
O'Connor, E., \& Ott, C.~D. 2011, ApJ, 730, 70.
\newblock \url{https://iopscience.iop.org/article/10.1088/0004-637X/730/2/70}

\bibitem[{Oey \& Garcia‐Segura(2004)}]{Oey2004b}
Oey, M.~S., \& Garcia‐Segura, G. 2004, ApJ, 613, 302.
\newblock \url{https://iopscience.iop.org/article/10.1086/421483}

\bibitem[{Oey {et~al.}(2017)Oey, Herrera, Silich, Reiter, James, Jaskot, \&
  Micheva}]{Oey2017}
Oey, M.~S., Herrera, C.~N., Silich, S., {et~al.} 2017, ApJ, 849, L1.
\newblock \url{http://dx.doi.org/10.3847/2041-8213/aa9215}

\bibitem[{{Oskinova} \& {Schaerer}(2022)}]{Oskinova2022}
{Oskinova}, L.~M., \& {Schaerer}, D. 2022, \aap, 661, A67

\bibitem[{{Patton} \& {Sukhbold}(2020)}]{Patton2020}
{Patton}, R.~A., \& {Sukhbold}, T. 2020, \mnras, 499, 2803

\bibitem[{{Petit} {et~al.}(2006){Petit}, {Drissen}, \& {Crowther}}]{Petit2006}
{Petit}, V., {Drissen}, L., \& {Crowther}, P.~A. 2006, \aj, 132, 1756

\bibitem[{Ramachandran {et~al.}(2019)Ramachandran, Hamann, Oskinova, Gallagher,
  Hainich, Shenar, Sander, Todt, \& Fulmer}]{Ramachandran2019}
Ramachandran, V., Hamann, W.-R., Oskinova, L.~M., {et~al.} 2019, A\&A, 625,
  A104.
\newblock
  \url{https://www.aanda.org/articles/aa/abs/2019/05/aa35365-19/aa35365-19.html}

\bibitem[{{Savage}(1984)}]{Savage1984}
{Savage}, B.~D. 1984, in NASA Conference Publication, Vol. 2349, NASA
  Conference Publication, ed. J.~M. {Mead}, R.~D. {Chapman}, \& Y.~{Kondo},
  3--16

\bibitem[{{Savage} {et~al.}(2001){Savage}, {Meade}, \& {Sembach}}]{Savage2001}
{Savage}, B.~D., {Meade}, M.~R., \& {Sembach}, K.~R. 2001, \apjs, 136, 631

\bibitem[{Sawant {et~al.}(2021)Sawant, Pellegrini, Oey, López-Hernández, \&
  Micheva}]{Sawant2021}
Sawant, A.~N., Pellegrini, E.~W., Oey, M.~S., López-Hernández, J., \&
  Micheva, G. 2021, ApJ, 923, 78.
\newblock \url{http://dx.doi.org/10.3847/1538-4357/ac2c85}

\bibitem[{Senchyna {et~al.}(2019)Senchyna, Stark, Chevallard, Charlot, Jones,
  \& Vidal-García}]{Senchyna2019}
Senchyna, P., Stark, D.~P., Chevallard, J., {et~al.} 2019, \mnras, 488, 3492

\bibitem[{{Senchyna} {et~al.}(2022){Senchyna}, {Stark}, {Charlot}, {Plat},
  {Chevallard}, {Chen}, {Jones}, {Sanders}, {Rudie}, {Cooper}, \&
  {Bruzual}}]{Senchyna2022}
{Senchyna}, P., {Stark}, D.~P., {Charlot}, S., {et~al.} 2022, \apj, 930, 105

\bibitem[{Silich \& Tenorio-Tagle(2018)}]{Silich2018a}
Silich, S., \& Tenorio-Tagle, G. 2018, \mnras, 478, 5112

\bibitem[{{Silich} {et~al.}(2020){Silich}, {Tenorio-Tagle},
  {Mart{\'\i}nez-Gonz{\'a}lez}, \& {Turner}}]{Silich2020}
{Silich}, S., {Tenorio-Tagle}, G., {Mart{\'\i}nez-Gonz{\'a}lez}, S., \&
  {Turner}, J. 2020, \mnras, 494, 97

\bibitem[{Silich {et~al.}(2004)Silich, Tenorio-Tagle, Rodríguez-González, \&
  Muñoz-Tuñón}]{Silich2004}
Silich, S., Tenorio-Tagle, G., Rodríguez-González, A., \& Muñoz-Tuñón, C.
  2004, ApJ, 610, 226

\bibitem[{Silich {et~al.}(2007)Silich, Tenorio‐Tagle, \&
  Munoz‐Tunon}]{Silich2007}
Silich, S., Tenorio‐Tagle, G., \& Munoz‐Tunon, C. 2007, ApJ, 669, 952.
\newblock \url{http://adsabs.harvard.edu/abs/2007ApJ...669..952S}

\bibitem[{{Silich} {et~al.}(2023){Silich}, {Turner}, {Mackey}, \&
  {Mart{\'\i}nez-Gonz{\'a}lez}}]{Silich2023}
{Silich}, S., {Turner}, J., {Mackey}, J., \& {Mart{\'\i}nez-Gonz{\'a}lez}, S.
  2023, \apjl, 944, L32

\bibitem[{Smith {et~al.}(2023)Smith, Oey, Hernandez, Ryon, Leitherer, Charlot,
  Bruzual, \& Al.}]{Smith2023}
Smith, L.~J., Oey, M.~S., Hernandez, S., {et~al.} 2023, ApJ, in press, arXiv:
  1805.09865.
\newblock \url{https://ui.adsabs.harvard.edu/abs/2023arXiv231003413S}

\bibitem[{Stark {et~al.}(2015)Stark, Walth, Charlot, Clément, Feltre, Gutkin,
  Richard, Mainali, Robertson, Siana, Tang, \& Schenker}]{Stark2015}
Stark, D.~P., Walth, G., Charlot, S., {et~al.} 2015, \mnras, 454, 1393

\bibitem[{Thuan {et~al.}(2014)Thuan, Bauer, \& Izotov}]{Thuan2014}
Thuan, T.~X., Bauer, F.~E., \& Izotov, Y.~I. 2014, \mnras, 441, 1841.
\newblock
  \url{http://academic.oup.com/mnras/article/441/2/1841/1078053/The-Xray-properties-of-the-cometary-blue-compact}

\bibitem[{Tolstoy {et~al.}(1995)Tolstoy, Saha, Hoessel, \&
  McQuade}]{Tolstoy1995}
Tolstoy, E., Saha, A., Hoessel, J.~G., \& McQuade, K. 1995, \aj, 110, 1640.
\newblock \url{http://adsabs.harvard.edu/cgi-bin/bib_query?1995AJ....110.1640T}

\bibitem[{{Turner} {et~al.}(2017){Turner}, {Consiglio}, {Beck}, {Goss}, {Ho},
  {Meier}, {Silich}, \& {Zhao}}]{Turner2017}
{Turner}, J.~L., {Consiglio}, S.~M., {Beck}, S.~C., {et~al.} 2017, \apj, 846,
  73

\bibitem[{Vink(2022)}]{Vink2022}
Vink, J.~S. 2022, \araa, 60, 203.
\newblock \url{https://doi.org/10.1146/annurev-astro-052920-094949}

\bibitem[{{Wang} \& {Yao}(2005)}]{Wang2005}
{Wang}, Q.~D., \& {Yao}, Y. 2005, in AIP Conference Series, Vol. 774, X-ray
  Diagnostics of Astrophysical Plasmas: Theory, Experiment, and Observation,
  ed. R.~{Smith}, 191--199

\bibitem[{{Yadav} {et~al.}(2017){Yadav}, {Mukherjee}, {Sharma}, \&
  {Nath}}]{Yadav2017}
{Yadav}, N., {Mukherjee}, D., {Sharma}, P., \& {Nath}, B.~B. 2017, \mnras, 465,
  1720

\bibitem[{{Yeh} \& {Matzner}(2012)}]{Yeh2012}
{Yeh}, S. C.~C., \& {Matzner}, C.~D. 2012, \apj, 757, 108

\end{thebibliography}
\end{document}